\documentclass[twocolumn,showpacs,aps,floatfix]{revtex4}

\usepackage{amsmath2000}
\usepackage{graphicx}% Include figure files
\usepackage{dcolumn}% Align table columns on decimal point
\usepackage{bm}% bold math

\begin{document}
\preprint{ND Atomic Theory 2002-3}
\title{Laser gas-discharge absorption measurements of the ratio of two transition rates in argon}
\author{I. M. Savukov}
 \email{isavukov@nd.edu}
 \homepage{http://www.nd.edu/~isavukov}
\author{H. G. Berry}
 \email{Berry.20@nd.edu}
 %\homepage{http://www.nd.edu/~johnson}
\affiliation{Department of Physics, 225 Nieuwland Science Hall\\
University of Notre Dame, Notre Dame, IN 46566}

\date{\today}
\begin{abstract}
The ratio of two line strengths at 922.7 nm and 978.7 nm of argon is measured in an argon 
pulsed discharge with the use of a single-mode Ti:Sapphire laser. 
The result 3.29(0.13) is in agreement with our theoretical prediction 3.23 and with a less 
accurate ratio 2.89(0.43) from the NIST database.   
\end{abstract}

\pacs{39.30.+w, 32.70.Cs, 32.80.-t}
\maketitle
\section{Introduction}

We have recently developed a theory for particle-hole states of closed-shell atoms
that is potentially able to predict energies and transition rates with high
precision. Initially, this theory has been tested in neon~\cite{neonepr,disser}.
For transitions between neon excited states the accuracy of
experiments is very good and experiments agree well with our theory. However, for
heavier noble gas atoms, experiments are often prone
to systematic errors and the precision is much lower. For example, many absolute transition
rates of argon were based on high-pressure arc plasmas and depended on the
assumption of local thermodynamic equilibrium of the upper state populations
which often was satisfied very approximately (see discussion in Ref.~\cite{Hiraby}). Lifetime
measurements using different techniques or even the same technique often yield quite
different results. In Ref.~\cite{argmes}, a unified set of atomic probabilities for
neutral argon is given. The data are obtained from extensive analysis of
existing experiments and from new measurements of emission in the broad
range 240-18000 nm. The data are included in the NIST data base. Errors of
order 10\% are quoted. Although our theory is able to reproduce NIST data,
the difference between experiment and theory often exceeds quoted in Ref.~\cite{argmes}
error bars. There are two possibilities: (i) the theory is less accurate and
needs to be improved, (ii) experimental error bars are larger and critical,
more precise measurements are needed. Therefore, the accurate measurements
presented in this paper are well motivated.

We describe a new experimental technique for precise measurements of
ratios of transition rates in gas discharges and detail an apparatus realizing this
technique. We demonstrate our measurement of the ratio of two argon
transitions and analyze systematic errors. Finally, we present a comparison
with our theory and other theories and experiments. 

\section{An experimental technique for accurate measurements of transition
rate ratios}

The essential goal of the experimental technique is to measure the relative
absorption rates for two different transitions that start from the same
level. In general, the populations of the lower and upper states are needed
for finding the cross-section $\sigma _{if}$ from the absorption coefficient 
$\alpha _{if}$, 
\begin{equation}
\alpha _{if}=\left[ N_{i}-(g_{i}/g_{f})N_{f}\right] \sigma _{if}
\end{equation}
(here $N_{i}$ and $N_{f}$ are densities of atoms in the initial and final
states, $g_{i}/g_{f}$ is the ratio statistical weights), but in some cases
the upper state population can be neglected. This happens when the initial
state is the ground or long-lived metastable state, and the final, upper
states are short-lived and relatively unpopulated. In gas discharges, at low
current densities, metastable levels are generally much more populated than
any other non-metastable excited levels. Experimentally, we also observed
states that have allowed decay modes. The $\left[ 3p_{3/2}^{-1}4s_{1/2}%
\right] _{1}$ and $\left[ 3p_{1/2}^{-1}4s_{1/2}\right] _{1}$ states,
surprisingly, were populated almost as much as metastables at pressure of a
few Torr. The observed lifetimes of these states are of the order of ten
microseconds. The principal explanation for such a slow decay is radiation
capture: many possibilities of emission and absorption of a photon occur
before the photon can leave the discharge cell. The observation of
long-lived species in discharges gives the possibility to use them as lower
states for absorption measurements. The other excited levels (not belonging
to the 4s group) quickly decay radiatively without radiation capture and are
effectively depopulated in a pulsed discharge as soon as the discharge
current disappears. Therefore, in upward transitions that start from the 4s
levels, the condition $N_{i}>>N_{f}$ is satisfied, and the absorption
becomes proportional just to the cross-section and the initial density.
Thus, by measuring the absorption for two transitions starting from the same
level, in our experiment $\left[ 3p_{1/2}^{-1}4s_{1/2}\right] _{1}$, we can
find the ratio of cross-sections from the ratio of absorptions, 
\begin{equation}
\frac{\sigma _{if_{1}}}{\sigma _{if_{2}}}=\frac{\alpha _{if_{1}}}{\alpha
_{if_{2}}}.
\end{equation}
For a Doppler broadened profile and for monochromatic laser light tuned to
the maximum of the profile, the ratio of cross-sections is equal to the
ratio of line strengths 
\begin{equation}
\frac{\sigma _{if_{1}}}{\sigma _{if_{2}}}=\frac{S_{if_{1}}}{S_{if_{2}}}=%
\frac{S_{1}}{S_{2}}\text{.}
\end{equation}
Therefore by measuring the ratio of absorptions for two transitions, we can
find the ratio of line strengths $S_{1}/S_{2}$, the important parameter that
can be compared with other experiments and with theory. The situation
described above is ideal for precise measurements of transition rate ratios:
it has been very closely reproduced in alkali vapor experiments, in which
for Cs an accuracy of better than 0.1\% is claimed by~\citet{csexp}. However,
in the case of noble gases, especially when we choose the starting level to
be a non-metastable state, a more complicated analysis is required.

In order to satisfy the condition $N_{i}>>N_{f}$ for the lowest excited states,
we use a pulsed rather than a DC discharge.  In a DC discharge this condition
is weaker, because radiatively decaying levels are constantly repopulated
by collisions with hot electrons and by decay of higher levels. 
It is especially true for radiation-capturing levels. A further advantage of a pulsed
discharge is the possibility for enhancing the signal by utilizing its time
dependence. For example slow fluctuations in laser intensity can be
excluded. In addition, the signal-to-noise ratio can be improved using
averaging (128-scan in our experiment) on a digital oscilloscope. Without
special efforts an absorption of order 0.1\% can be easily detected. Such
high sensitivity is useful for fast and precise tuning of the laser to
centers of Doppler profiles which is needed if only one laser is available.
The long-term stability of the pulsed discharge is also better: the pulsed
discharge has substantially smaller sputtering and thermal effects, since
the total charge from the discharge current is much smaller for a pulsed
discharge.

The principal disadvantage of the pulsed discharge is the more complicated
analysis of systematic errors, basically owing to the finite-time response
of the detection system. The various factors that can influence the
measurement of the ratio in pulsed discharges need careful analysis: we will
consider systematic errors that arise from large absorption and
inhomogeneity. When two transitions differ by a factor of at least
three, a larger absorption is
recommended for better signal-to-noise ratio. In such a case the linear
relation between absorption and line strength may no longer hold, and
instead Beer's law must be used to derive the more general equation: 
\begin{equation}
\frac{\ln (I_{1}/I_{10})}{\ln (I_{2}/I_{20})}=\frac{S_{1}}{S_{2}}.
\label{ratiolog}
\end{equation}
Here $I_{1}/I_{10}$ and $I_{2}/I_{20}$ are the reductions in the intensity
after passing the absorber for the first and the second transitions,
respectively. This nonlinear equation can be the cause of several systematic
errors.

One effect arises from a slow response of the detection system. This results
in a reduction of actual height of the peak by an unknown coefficient $\eta $%
: 
\begin{equation}
\frac{\ln (\eta I_{1}/I_{10})}{\ln (\eta I_{2}/I_{20})}=\frac{S_{1}}{S_{2}}.
\end{equation}
If the equations were linear, this coefficient would drop out from the
ratio, but for large and significantly different absorptions for the two
transitions, it will cause a systematic error in the ratio. The other effect
arises from inhomogeneity. So far we assumed that the absorption medium is
homogeneous, but in gas discharges this is far from being true. The
discharge produces densities of excited atoms with some transverse and
longitudinal distributions. The longitudinal distribution does not cause any
problem, since for two transitions the ratio at each small length of path
stays the same, and eventually the ratio of logarithms of intensities in
Eq.~(\ref{ratiolog}) will remain the same. The transverse distribution,
however, will cause a deviation from the simple Beer's law and will
require the analysis of this distribution.

For a cylindrical discharge cell, which we use in our experiment, if $r$ is
the distance from the axis, an integration over $r$ is needed, 
\begin{equation}
I_{out}=\int I_{in}(r)\exp \left[ -\alpha (r)\right] 2\pi rdr\text{,}
\end{equation}
where $I_{in}(r)$ is the input laser beam intensity, $\alpha (r)$ is the
absorption coefficient, and $I_{out}$ is integrated over a detector area
output intensity after an inhomogeneous discharge medium. For simplicity in
this equation the change in absorption along the laser beam is
ignored. It is difficult to find the spatial distribution accurately,
although in some simple cases it can be predicted. In the case of a
diffusion controlled cylindrical discharge, the distribution for the
electron density $n_{e}$ and for excited state densities, has approximately
the form of a Bessel function, 
\begin{equation}
\alpha (r)\sim n_{e}\approx J_{0}(2.4r/R)\cos (\pi z/L).
\end{equation}
The expression for the electron density $n_{e}$ is taken from Ref.~\cite
{gasdischarge}, p. 67. Using the series expansion of the Bessel function for
small argument, $J_{0}(2.4r/R)\approx 1-\frac{1}{4}(2.4r/R)^{2}$ we can show
that in our experiment with $R=1.25$ cm and $r=0.2$ cm, the
maximum change in the absorption over the beam radius $r$ is 3.7\% and the
change in the integral for small $\alpha (r)$ is only 1.8\%. Therefore the
effect of the distribution is small. Moreover, if the absorption is not very
large, the exponential can be expanded, and the linear dependence of the
signal on the absorption coefficient can reduce further the effect of the
distribution.
%%%%%%%%%%%%%%%%%%%
If we examine NIST transition data~\cite{nist:01} for neutral argon, we find many
transitions in the range of tunable lasers such as a Ti:Sapphire laser,
which can be used for the absorption experiments. However, in practice there
are some restrictions: first, the pair of transitions has to start from the
same metastable levels or radiation-capturing levels such as mentioned
above. Second, the range is greatly restricted by the tunability of the
laser: for example, our Ti:Sapphire laser has several mirror sets for
several ranges of wavelengths. Because the change of mirrors takes many
hours to complete, only one mirror set can be used for measurements of
ratios at one time. As a result, for our currently installed mirror set,
only two pairs of transitions were appropriate. However, our laser could not
be tuned for one transition, and we ended up able to make measurements for
only one pair of transitions: one transition at a wavelength 9227~\AA\ and
the other at 9787~\AA . 

\section{Experimental arrangement}

\begin{figure}[tbp]
\centerline{\includegraphics*[scale=0.5]{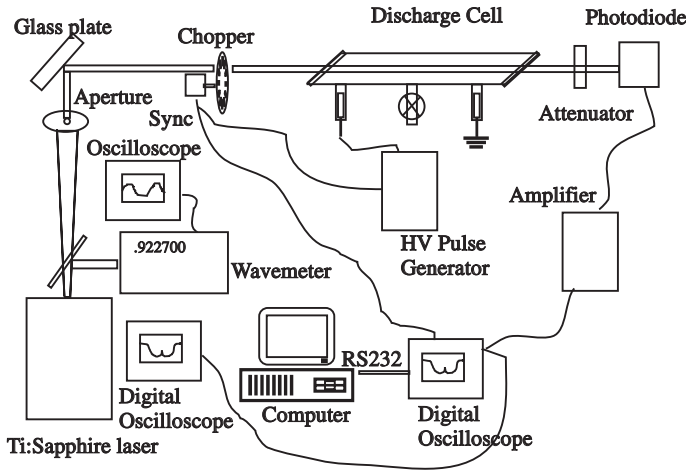}}
\caption{An experimental setup for conducting measurements of absorption of
laser radiation in pulsed discharges}
\label{figa1}
\end{figure}
The optical and electrical diagrams of the experimental arrangement are shown in
Fig.~\ref{figa1}. The monochromatic 1W laser beam generated by the
Ti:Sapphire laser is expanded by traveling 8 meters between two optical
tables to the size of about 5 cm diameter. A variable aperture reduces both
the beam's size and its power. To avoid saturation of the transition, the
beam intensity is further reduced using a 2\% reflection off a glass plate.
A chopper periodically interrupts the beam to give a 100\% reference
absorption. The beam is then directed through the discharge cell where it
undergoes time-dependent absorption in the discharge initiated by HV pulses
from the HV pulse generator. The optical path of the laser beam is
terminated on the photodiode. The attenuator in front of the screened
photodiode compartment is used to avoid any saturation effects in the
detection system. The signal from the photodetector is amplified and sent to
the digital oscilloscope for measurement of the absorption.

The laser is a commercial single frequency MBR-110 Ti:Sapphire laser, pumped
with 7-15 W of CW Coherent argon ion laser. The MBR-110 laser can be tuned
in a broad range, 700-1060 nm, but to cover all this range 6 different
mirror combinations have to be installed.
% Six mirror groups provide the
%following spectral ranges of operations: 700-780 nm, 731- 872 nm, 854-906
%nm, 873-964 nm, 957-993 nm, 970-1060 nm. In this experiment the 957-993 nm
%mirror set was used. In this particular range, 16 argon transitions can be
%found in the NIST database, but only one pair, at 922.7 nm and at 978.7 nm,
%were suitable for precise measurements. Another pair, though in the range of
%our installed mirror set, was not suitable because laser operation at the
%936 nm transition wavelength was not stable. The cause of instability is
%unknown: it may be the presence of water vapor.

The discharge cell is a quartz tube of diameter 2.5 cm and a length of 30 cm.
Two standard electrodes were fused symmetrically in side-tubes
near each end. The electrodes have large area to reduce unwanted sputtering
from the cathode which can reduce the lifetime of metastables significantly
and affect the population distribution in the discharge as we observed in
detail in helium. Moreover, for the same purpose, an external ballast
volume was added to the cell and a short-pulsed discharge was used instead
of a DC discharge.

A high voltage pulse generator produces  high-voltage (more
than 10 kV to allow fast discharge breakdown) pulses each of a few microseconds. 
The circuit is based
on a power transistor in the switching mode and high-voltage transformer. A
rotating chopper is inserted into the optical path to produce a 100\%
normalization signal.
%Other elements such as mirrors, a glass plate, an
%aperture, and a beam attenuator, serve to direct the laser beam and to
%reduce the intensity and the size of interacting laser beam.

The detection system consists of a photodiode, an amplifier, and a digital
oscilloscope. There is no special requirement on those elements except that
the detection system should provide a linear response and the distortion of
the waveform of the signal should be minimal. The first requirement is
completely fulfilled if the laser intensity does not exceed the saturation
limit for a particular diode. The second requirement is approximately
satisfied: small distortion of the signal has been observed. This effect is
analyzed in detail in the next section.

\section{Measurement and error analysis}

\begin{figure}[b]
\centerline{\includegraphics*[scale=0.45]{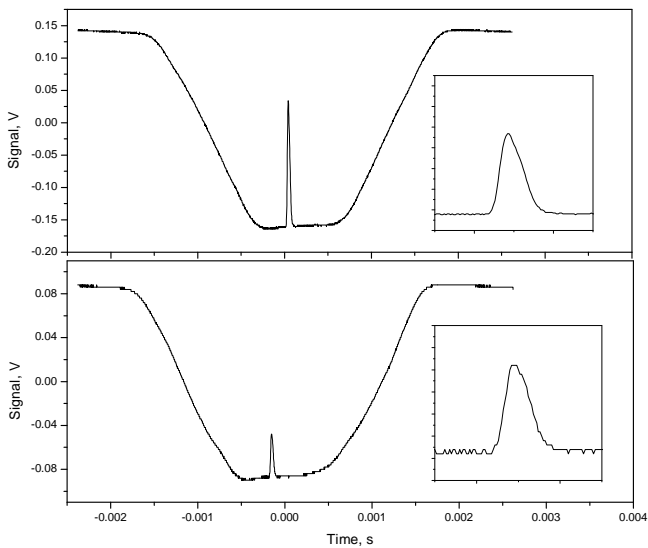}}
\caption{Absorption signals in a pulsed discharge with chopper
modulation for normalization. Absorption of two transitions is measured.
Negative voltages occur because the dc component of the signal is removed.
The insets show that after rescaling the absorption peaks have the same
shapes }
\label{figa3}
\end{figure}

To find the ratio of line strengths, we measure the peak absorption for two
transitions and use the formula 
\begin{equation*}
\frac{\ln (1-V_{1}/V_{10})}{\ln (1-V_{2}/V_{20})}=\frac{S_{1}}{S_{2}}\text{.}
\end{equation*}
Here $V_{1}$ and $V_{10}$ are the voltages proportional to the maximum of a
pulsed discharge absorption and to the chopper normalization absorption for
the transition at $\lambda =922.7$ nm shown in Fig.\ref{figa3}. Similar
voltages for the second transition at $\lambda =978.7$ nm also shown in Fig.\ref
{figa3}. are $V_{2}$ and $V_{20}$.

Repeating measurements of absorption several times, we have obtained the
ratio $3.29\pm 0.04$. A small error here indicates the excellent statistics.
However, the result can still have systematic errors which need to be
estimated and if possible be excluded.

The first type of systematic error can occur from a slow drift in discharge
conditions. Though our pulsed discharge is very stable, to account for any
slow drift, we alternate measurements of the absorptions for the two transitions. 

The second possible systematic error can arise from the uncertainty in the
wavelength when the laser is manually tuned to the resonances of the
transitions by monitoring absorption peaks similar to shown in Fig.~\ref
{figa3}. If $\varepsilon _{1}$ and $\varepsilon _{2}$ are relative average
shifts owing to tuning uncertainty, then the absolute systematic shift in
the ratio $\sigma \left( \varepsilon _{1},\varepsilon _{2}\right) $ is

\begin{equation}
\sigma \left( \varepsilon _{1},\varepsilon _{2}\right) =\frac{\ln \left[
(1-\left( 1-\varepsilon _{1}\right) V_{1}/V_{10})\right] }{\ln \left[ \left(
1-\left( 1-\varepsilon _{2}\right) V_{2}/V_{20}\right) \right] }-\frac{S_{1}%
}{S_{2}}\approx (\varepsilon _{2}-\varepsilon _{1})\text{,}
\end{equation}
where this equation was simplified for small absorption. The tuning error is
estimated to be  about 1.5\%.

The combined effect of the nonlinearity of absorption and finite-time
response of the detection system leads to another systematic error.
Suppose $\varepsilon $ is the reduction in the peak height because of a slow
time response. Then we can write for the $S$-ratio 
\begin{equation*}
\frac{\ln \left[ 1-\left( 1-\varepsilon \right) V_{1}/V_{10}\right] }{\ln %
\left[ 1-\left( 1-\varepsilon \right) V_{2}/V_{20}\right] }=\frac{S_{1}}{%
S_{2}}+\sigma 
\end{equation*}
Comparing measurements of ratios at different discharge currents and, hence
absorption,    we find that the time responce error is about $4\%$. This
error gives a systematic increase in the ratio by 4\%, but the previous
error due to tuning decreases the ratio slightly by 1.5\% so that the
overall systematic error is smaller. Finally, taking into account mentioned
above statistical and systematic errors, we obtain the accuracy of the
measurement of the ratio about 4\%. The final value of our ratio is $3.29\pm
0.13$. 

\section{Comparison with CI+MBPT theory and the other measurements and
calculations}
\begin{table}[tbp]
\caption{The ratio of the two line strengths at 922.7 nm and 978.7 nm}
\label{argtb}
\begin{center}
\begin{tabular}{ll}
\hline\hline
Source  &   Ratio \\
\hline
Experiment &  \\
~This work          & 3.29$\pm$0.13 \\
~\citet{argmes}     & 2.9$\pm$0.4 \\
\hline
Theory  &   \\
~\citet{disser}     &  3.23  \\
~\citet{Lilly}      &  3.51 \\
~\citet{gvb}        &  4.62 \\
\hline
\end{tabular}
\end{center}
\end{table}
The comparison of the ratio of line strengths is given in Table~\ref{argtb}. Our measured ratio of line 
strengths $3.29\pm 0.13$ is in excellent
agreement with the value 3.23 obtained from our mixed
configuration-interaction and perturbation-theory calculations~\cite{disser}. Both
measured and theoretical ratios agree also with the ratio 2.89$\pm $15\%
calculated from the experimental transition rates of Ref.~\cite{argmes}. Our
results are close to semiempirical calculations by~\citet{Lilly}.
However, the ratio 4.62 from semiempirical calculations of Ref.~\cite{gvb} based
on a similar principle as calculations in Ref.~\cite{Lilly} disagree substantially
with our result. One obvious problem with semiempirical approaches is that
energies can not define completely the wave functions which are
very sensitive to the accuracy of calculations. In addition, transition
amplitudes have additional corrections which are not accounted by the
effective Hamiltonian. For example, random-phase approximation (RPA)
corrections are due to shielding by the atom of an external photon field.
Those corrections are substantial in Ar and heavier noble-gas atoms and must
be included to reach a good precision.

\section{Conclusion}

We have measured an accurate ratio of line strengths for two argon
transitions. For this measurement we developed a new experimental technique
based on a pulsed discharge and a laser absorption measurement. This
technique can be applied for measurements of  other ratios in noble gas
atoms. The agreement of our measurement and calculations is excellent. 
%We also agree with the latest best experimental values and with a successful
%semiempirical theory of Lilly. 

%\begin{acknowledgments}
%I. M. S. was supported in part by National
%Science Foundation Grant No.\ PHY-01-39928. 
%\end{acknowledgments}

%\bibliography{disser}

\end{document}